\documentclass[a4paper]{article}
\usepackage[T1]{fontenc}
\usepackage{times}
\usepackage{soul}
\usepackage{url}
\usepackage[hidelinks]{hyperref}
\usepackage[utf8]{inputenc}
\usepackage[small]{caption}
\usepackage{graphicx}
\usepackage{amsmath}
\usepackage{amsthm}
\usepackage{booktabs}
\usepackage{algorithm}
\usepackage{algorithmic}
\usepackage{multirow}
\usepackage{threeparttable}
\usepackage{amsfonts}
\usepackage{float}
\usepackage{url}
\usepackage{bm}
\usepackage{xcolor}
\usepackage{subfigure}
\usepackage{amssymb}
\usepackage{footmisc} 

\usepackage{INTERSPEECH2022}

\title{Cross-speaker Emotion Transfer Based On Prosody Compensation for End-to-End Speech Synthesis}
\name{Tao Li$^1$, Xinsheng Wang$^2$, Qicong Xie$^1$, Zhichao~Wang$^1$, Mingqi Jiang$^3$, Lei Xie$^1$$^{\ast}$
\thanks{*Corresponding author. This work was supported by the National Key R \& D Program of China (2020AAA0108600).}}

\address{
  $^1$Audio, Speech and Language Processing Group (ASLP@NPU), School of Computer Science,\\ Northwestern Polytechnical University, Xi’an, China \\
  $^2$School of Software Engineering, Xi’an Jiaotong University, Xi’an, China\\
  $^3$Mobvoi AI Lab, Suzhou, China}
\email{taoli@npu-aslp.org, w.xinshawn@gmail.com,\{xieqicong, zcwang\_aslp\}@mail.nwpu.edu.cn, mingqi.jiang@mobvoi.com,lxie@nwpu.edu.cn}

\begin{document}
\maketitle
%
\begin{abstract}
Cross-speaker emotion transfer speech synthesis aims to synthesize emotional speech for a target speaker by transferring the emotion from reference speech recorded by another (source) speaker. In this task, extracting speaker-independent emotion embedding from reference speech plays an important role. However, the emotional information conveyed by such emotion embedding tends to be weakened in the process to squeeze out the source speaker's timbre information. In response to this problem, a prosody compensation module (PCM) is proposed in this paper to compensate for the emotional information loss. Specifically, the PCM tries to obtain speaker-independent emotional information from the intermediate feature of a pre-trained ASR model. To this end, a prosody compensation encoder with global context (GC) blocks is introduced to obtain global emotional information from the ASR model's intermediate feature. Experiments demonstrate that the proposed PCM can effectively compensate the emotion embedding for the emotional information loss, and meanwhile maintain the timbre of the target speaker. Comparisons with state-of-the-art models show that our proposed method presents obvious superiority on the cross-speaker emotion transfer task.
\end{abstract}

\section{Introduction}
\label{sec:intro}
While the development of attention-based sequence-to-sequence (seq2seq) neural models~\cite{Bahdanau2015NeuralMT,Sutskever2014SequenceTS} has brought dramatic improvement for the naturalness of synthetic speech~\cite{wang2017tacotron,Ren2019FastSpeechFR,yu2019durian}, there is still a big gap between synthetic speech and real human speech. Specifically, beyond the naturalness, natural speech produced by human beings is diverse in terms of emotions, which is important paralinguistic information conveyed by human speech. Therefore, how to present appropriate emotions in synthetic speech is vital in immersive human-computer interaction systems, and thus has been drawn much attention recently~\cite{Choi2019MultispeakerEA,Se2020Emotional,Wang2018Style, Zhang2019LearningLR,Bian2019Multi,Pan2021CrossspeakerST,Li2021ControllableET}. 

A simple case of emotional speech synthesis is the \textit{same-speaker}~\cite{Lee2017Emotional,Li2018EmphaticSG,Li2021ControllableET} scenario, in which the speaker identity of the training data is the same as that of synthesized speech. The limitation of this scenario is obvious, i.e., it can only be used for a specific target speaker that has enough desired emotion recordings. Emotional voice conversion~\cite{Zhou2020TransformingSA,Zhou2021EmotionalVC,Cao2020NonparallelES,Choi2021SequencetoSequenceEV} is one related task, which aims to convert the emotional state of the utterance from one to another while preserving the linguistic information and speaker identity. In contrast, the \textit{cross-speaker} emotional speech synthesis tries to transfer the emotion from a source speaker who has recorded emotional speech to another (target) speaker, enabling the TTS model to produce emotional speech with the timbre of a speaker who has no emotional speech recordings at all. Reference-based style transfer is the most popular strategy for this cross-speaker case~\cite{Gibiansky2017DeepV2,Skerry2018Towards,Wang2018Style,Zhang2019LearningLR,Bian2019Multi}. Reference encoder~\cite{Skerry2018Towards}, global style tokens (GST)~\cite{Wang2018Style,Bian2019Multi,Kwon2019EmotionalSS}, and variational autoencoder (VAE)~\cite{Zhang2019LearningLR,Kulkarni2020IMPROVINGLR} are commonly used methods to extract the emotion embedding from the reference with the desired emotion. In these methods, \textit{disentangling} the speaker information from the emotion embedding~\cite{Bian2019Multi,Whitehill2020MultiReferenceNT,Li2021ControllableCE} is important. Otherwise, the speaker information retained in the embedding could contaminate the target timbre.

However, it is challenging to prevent emotional information of the emotion embedding being weakened in the process of removing speaker information, because both the emotion and timbre are related to the prosody, making those reference-based methods have to make a trade-off between the emotion transfer quality and speaker similarity~\cite{Karlapati2020CopyCatMF}. Specifically, the reference embedding with enough emotion information tends to result in the \textit{speaker leakage}, while further disentangling the speaker information for the embedding could make the emotional expressiveness become weaker in synthetic speech. To face this challenge, in this paper we propose to compensate the emotion embedding for the emotional information loss that caused by the speaker information disentangling, to improve the emotional expressiveness of synthetic speech.
 
\begin{figure}[h]	
	\centering
	\includegraphics[width=1\linewidth]{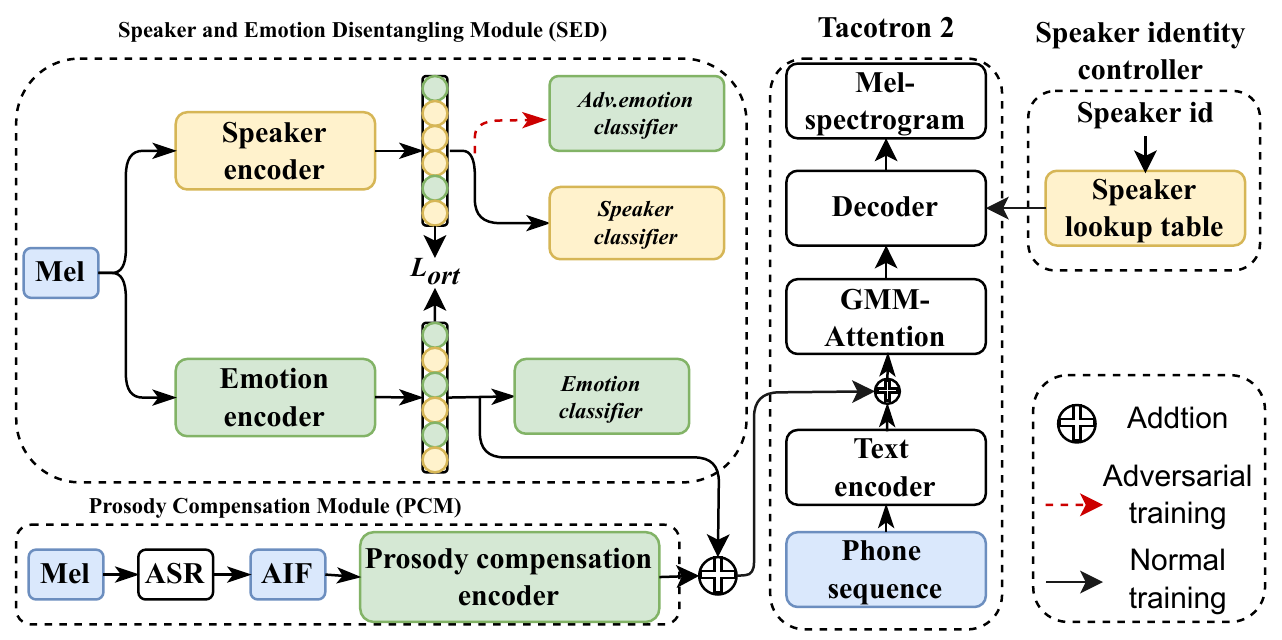}
	\vspace{-0.25cm}
	\captionsetup{belowskip=-8pt}
	\caption{General architecture of the proposed cross-speaker model.}
	\label{fig:frame1}
\vspace{-0.4cm} 
\end{figure}

Taking inspiration from~\cite{Wang2021AccentAS,Wang2021EnrichingSS} that the hidden representation produced by a pre-trained automatic speech recognition (ASR) model retains prosody information but no obvious speaker information, a prosody compensation module (PCM), which takes the ASR model's intermediate feature (AIF) of reference audio as input (as shown in the lower-left corner of Fig.~\ref{fig:frame1}), is proposed to compensate the emotion information. The proposed cross-speaker emotional speech synthesis model with prosody compensation, referred to as CSPC, is a Tacotron2~\cite{shen2018natural} based framework with extended speaker disentangling module (SDM), speaker identity controller, and PCM. To be specific, the SDM is to obtain disentangled speaker-independent emotion embeddings from the reference spectrum, and the PCM is to obtain extra emotion information from AIF to compensate for the information loss in the emotion embedding caused by the disentangling. To effectively extract global prosody information from AIF, a prosody compensation encoder assisted by global context~\cite{Cao2019GCNetNN} (GC) blocks is introduced. Extensive experiments have demonstrated the effectiveness of this proposed prosody compensation method on improving the emotional expressive and meanwhile maintaining the voice of the target speaker, and also indicated the effectiveness of the proposed prosody compensation encoder. 

In addition to the reference-based method, the most recent work~\cite{Pan2021CrossspeakerST} tried to synthesize diverse emotional speech with a label-based approach. To this end, a content-aware prosody prediction module is proposed to predict the prosody from input text with the explicit emotion label. It has shown that this label-based method with extra text-based prosody prediction can achieve comparable performance to the reference-based method. In this paper, in addition to several state-of-the-art reference-based methods, this label-based method will also be compared.

\vspace{-0.4cm}  

\vspace{-0.15cm}
\section{Proposed Model}
\label{sec:model}

The architecture of the proposed model is illustrated in Fig.~\ref{fig:frame1}, which shares the skeleton of the modified Tacotron2~\cite{Yang2019ImprovingME,Li2021ControllableET} while with proposed SDM to obtain speaker-independent emotion embeddings by squeezing out the source timbre information, and PCM to compensate emotion information. The target speaker’s identity is provided by the speaker identity controller. 

\subsection{Speaker Disentangling Module (SDM)}
\label{sec:mi}

In the reference-based \textit{cross-speaker} emotion transfer speech synthesis method, the emotion embedding obtained from reference audio should be emotionally salient and free of speaker information. To this end, following the basic idea of the speaker disentangling module in \cite{Li2021ControllableCE}, the emotion encoder trained with SDM is restrained by an emotion classification constraint to make the obtained emotion embedding discriminative in terms of emotion categories. The corresponding loss is referred to as ${\cal L}_{emo}$. 

To keep source speaker information out of the emotion embedding, the emotion embedding is made orthogonal to the speaker embedding with an orthogonality loss ${\cal L}_{ort}$~\cite{Li2021ControllableCE}, which is defined as
\begin{equation}
    {{\cal L}_{ort}} = \sum\limits_{i = 1}^n {\left\| {s_i^T{e_i}} \right\|_F^2} 
\label{eq:orth_loss}
\end{equation}
where $\left \| \cdot \right \|_{F}$ is the Frobenius norm and the $e_i$ is emotion embedding. The speaker embedding $s_i$ is obtained by the speaker encoder in SDM, which is trained with the speaker classification loss ${\cal L}_{spk}$ and an emotion classification loss with gradient inversion ${\cal L}_{adv\_emo}$.

The emotion encoder and speaker encoder share the same input and architecture that consist of six 2D convolutional layers and a GRU layer, and only the last GRU state is taken as the global feature that works as the input of an FC layer to generate the 256-dim embedding.

\vspace{-0.2cm}
\subsection{Prosody Compensation Module (PCM)}
\label{sec:pc}
Due to the entanglement of emotional information and speaker information in speech, the emotional information conveyed by the emotion embedding would be weakened after disentangling the speaker-related information, which could lead to insipid expressiveness in synthetic speech. Inspired by the research~\cite{Wang2021AccentAS,Wang2021EnrichingSS} which shows that the ASR's intermediate feature (AIF) retains the speaker-independent emotional information, PCM is proposed here to compensate the emotion embedding for the emotional information loss using AIF. As illustrated in Fig.~\ref{fig:frame1}, the AIF extracted by the encoder of a pre-trained ASR model is used to obtain a global feature via the prosody compensation encoder.

\begin{figure}[h]	
	\centering
	\includegraphics[width=1\linewidth]{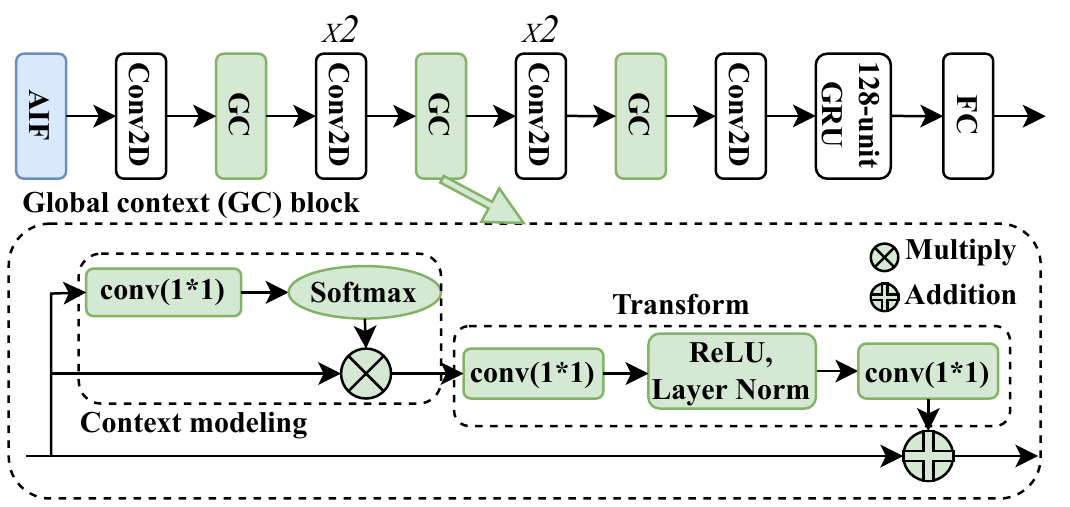}
	\vspace{-0.2cm}
	\captionsetup{belowskip=-8pt}
	\caption{The architecture of prosody compensation encoder}
	\label{fig:frame2}
\end{figure}

While the contextual information contained by the sequence is important for an ASR model, at a specific time step, the prediction of a word or phone is more related to the local context sequence rather than the entire sequence. In contrast, the emotion conveyed by an entire utterance is global information in general. Therefore, how to extract this global information from relatively scattered ASR model's intermediate representations is important. To this end, the global context~\cite{Cao2019GCNetNN} (GC) block, which shows promising ability on obtaining global information, is used in the prosody compensation encoder. As shown in Fig.~\ref{fig:frame2}, this prosody compensation encoder is similar to the emotion encoder in SDM. The difference is that three GC blocks are introduced in this encoder. The GC block~\cite{Cao2019GCNetNN} consists of global attention pooling, bottleneck transform, and broadcast element-wise addition for context modeling, channel-wise dependency capturing, and feature fusion, respectively, which help the encoder to capture long-range dependency. The prosody compensation embedding produced by this prosody compensation encoder is added to the emotion embedding, which is then added with the tacotron encoder output to work as the conditioning variable for providing emotion information.

\vspace{-0.2cm}
\subsection{Training and Generation}
\vspace{-0.1cm}
\label{sec:train}

Here, a Transformer-based ASR model, which is implemented with the WENET\footnote{The model is available at \url{https://github.com/wenet-e2e/wenet}}~\cite{yao2021wenet} toolkit and pre-trained on multi-Chinese corpora containing 2825 hours, is utilized. The encoder of this ASR model produces 256-dim AIF. Parameters of SDM and prosody compensation encoder are optimized together with the skeleton of Tacotron2. The loss function of this skeleton is the same as the objective function of Tacotron2 \cite{shen2018natural}, which is denoted as ${{\cal L}_{taco}}$. The final objective function to train the proposed model is given by:
\begin{equation}
\setlength{\abovedisplayskip}{3pt}
\setlength{\belowdisplayskip}{3pt}
    {\cal L} = {{\cal L}_{taco}} + {{\cal L}_{emo}} + {{\cal L}_{spk}} + {\cal L}_{adv\_emo}  + \alpha {\cal L}_{ort} ,
\end{equation}
where $\alpha = 0.1 $ is hyper-parameters of ${\cal L}_{ort}$.

For the speaker information, we specify the timbre for the model by a trainable speaker look-up table~\cite{Skerry2018Towards} during the training and inference stage, which takes the speaker ID as input and generates a 128-dimensional speaker representation to concatenate with the decoder's input in frame-level. The emotion information is provided by the emotion encoder and prosody compensation module, both of which are with the source speaker's emotional speech as input.

\vspace{-0.2cm}
\section{EXPERIMENTS}
\label{sec:majhead}
\vspace{-0.1cm}

\subsection{Experimental Setup}
\label{ssec:subhead}
\vspace{-0.2cm}


To achieve emotion transfer across speakers, a database that consists of at least two speakers is required, in which one is the source speaker to provide reference emotional speech and another one is the target speaker who only has neutral emotion speech. In this work, the source speaker comes from an individual adult female emotional corpus, referred to as \textit{DB\_6}. \textit{DB\_6} consists of 5,000 neutral utterances and 2,000 utterances for each emotion category (surprise, happy, sadness, angry, disgust, and fear). The corpus providing the target speaker is \textit{DB\_2}, which is another individual adult female corpus that contains about 5,000 samples recorded in a reading style with neutral emotion. Timbres of the source speaker and the target speaker are distinctly different.

During training, all speech samples are down-sampled to 16 kHz, and are then represented by Mel-spectrograms obtained with $50$ ms frame length and $12.5$ ms frame-shift. WaveRNN~\cite{Kalchbrenner2018EfficientNA} is adopted as the neural vocoder to reconstruct waveform from the predicted Mel-spectrograms during inference. Twenty sentences from the target speaker's test set were randomly selected to synthesize speech for each emotion category, resulting in 120 testing samples for each evaluated method.

\vspace{-0.2cm}
\subsubsection{Evaluation metrics}
\label{sssec:subsubhead}

A good synthetic sample in the emotion transfer task should present the emotion of the reference audio and meanwhile keep the timbre of the target speaker. Here, typical MOS (Mean Opinion Score) tests are conducted to evaluate the emotion similarity and speaker similarity subjectively. In the emotion similarity MOS test, participants are asked to rate synthetic speech based on how well this synthetic speech conveys the same emotion as that conveyed by reference audio, while the speaker similarity MOS test is to ask participants to rate synthetic speech based on how well the synthesized sample sound like uttered by the target speaker. The rating score in both MOS test ranges from \textit{1} to \textit{5} indicating the performance from \textit{bad} to \textit{great}. Besides, another subjective test, named as AB preference test, is adopted in the ablation study to compare samples synthesized by two models. In this AB test, participants are asked to choose one from two samples according to a certain requirement. For each subjective evaluation test, twenty Chinese native speakers participated.

\begin{table}[h]
 \caption{Comparison with state-of-the-art methods in terms of emotion similarity MOS with confidence intervals of 95\%.}
 \label{tab:compare_emomos}

\setlength{\tabcolsep}{1.3mm}
 \centering
\begin{tabular}{l|l|l|l|l}
\toprule
\multicolumn{1}{c|}{Method} & \multicolumn{1}{c}{Multi-R~\cite{Bian2019Multi}} & \multicolumn{1}{c}{CSET~\cite{Li2021ControllableCE}} & \multicolumn{1}{c}{PB~\cite{Pan2021CrossspeakerST}} & \multicolumn{1}{c}{CSPC} \\ \midrule
    surprise   &3.52$\pm$0.06  &3.73$\pm$0.05  &3.63$\pm$0.04 &\bf{3.92}$\pm$0.06 \\
    happy     &3.73$\pm$0.05  &3.85$\pm$0.06  &3.76$\pm$0.05  &\bf{4.01}$\pm$0.05 \\
    sadness   &3.30$\pm$0.04  &\bf{4.01}$\pm$0.04  &3.35$\pm$0.06  &\bf{4.07}$\pm$0.05 \\
    angry     &3.27$\pm$0.06  &3.69$\pm$0.05  &3.39$\pm$0.04  &\bf{3.85}$\pm$0.05 \\
    disgust   &3.21$\pm$0.06  &3.53$\pm$0.04  &3.31$\pm$0.05  &\bf{3.81}$\pm$0.06 \\
    fear      &3.25$\pm$0.04  &3.78$\pm$0.05  &3.47$\pm$0.05  &\bf{3.98}$\pm$0.05 \\\midrule
    average  &3.38$\pm$0.5 &3.76$\pm$0.05  &3.49$\pm$0.04 &\bf{3.89}$\pm$0.05 \\
  \bottomrule
\end{tabular}
\vspace{-0.4cm} 
\end{table}

\vspace{-0.3cm}
\section{Experimental Results}
\label{sec:page}

\subsection{Comparison with other methods}
\label{sc:comparsion}

To evaluate the performance of the proposed method on the \textit{cross-speaker} emotion transfer TTS task, we first compare our method with several state-of-the-art methods~\cite{Bian2019Multi,Li2021ControllableCE,Pan2021CrossspeakerST}. Among those compared methods, multi-reference Tacotron (Multi-R)~\cite{Bian2019Multi} and cross-speaker emotion transfer (CSET)~\cite{Li2021ControllableCE} are reference-based method, and prosody bottleneck (PB)~\cite{Pan2021CrossspeakerST} is a recently proposed label-based method.

Results of emotion similarity MOS test achieved by different models are shown in Table~\ref{tab:compare_emomos}, in which higher emotion similarity MOS value means better emotion transfer performance. As shown in this table, the proposed CSPC achieves the best emotion similarity in terms of all emotion categories. Besides, compared with other methods, the proposed CSPC presents a more stable performance in terms of different emotion categories. Specifically, the relative difference between the highest and lowest scores achieved by CSPC in different emotions is 6.8\%, while this differences of Multi-R, CSET, and PB are 16.2\%, 13.6\%, and 13.5\% respectively. 

\begin{table}[h]
 \caption{Comparison with sate-of-the-art methods in terms of speaker similarity MOS with confidence intervals of 95\%.}
 \label{tab:compare_spkmos}
\setlength{\tabcolsep}{1.3mm}
 \centering
\begin{tabular}{l|l|l|l|l}
\toprule
\multicolumn{1}{c|}{Method} & \multicolumn{1}{c}{Multi-R~\cite{Bian2019Multi}} & \multicolumn{1}{c}{CSET~\cite{Li2021ControllableCE}} & \multicolumn{1}{c}{PB~\cite{Pan2021CrossspeakerST}} & \multicolumn{1}{c}{CSPC} \\ \midrule
    surprise  &\bf{4.08}$\pm$0.06  &3.95$\pm$0.05 &\bf{4.05}$\pm$0.05 &3.97$\pm$0.04 \\
    happy     &\bf{4.15}$\pm$0.07  &3.83$\pm$0.07 &3.90$\pm$0.07  &3.85$\pm$0.04 \\
    sadness   &\bf{4.02}$\pm$0.05  &3.82$\pm$0.07 &3.86$\pm$0.08  &3.84$\pm$0.05 \\
    angry     &\bf{4.12}$\pm$0.05  &3.89$\pm$0.06 &3.93$\pm$0.06  &3.92$\pm$0.06 \\
    disgust   &4.0$\pm$0.06   &\bf{4.03}$\pm$0.04 &3.99$\pm$0.07  &\bf{4.01}$\pm$0.06 \\
    fear      &\bf{4.15}$\pm$0.07  &3.98$\pm$0.06 &4.06$\pm$0.04  &4.05$\pm$0.05 \\\midrule
    average  &\bf{4.08}$\pm$0.04  &3.91$\pm$0.06  &3.97$\pm$0.05 &3.94$\pm$0.05 \\
  \bottomrule
\end{tabular}
\end{table}

In addition to transferring the emotion from reference to synthesized speech, the timbre of the target speaker should be kept in the synthesized speech. Table~\ref{tab:compare_spkmos} shows the performance of different models on keeping the target timbre in synthetic speech, which is evaluated by the speaker similarity MOS test. Higher speaker similarity means better timbre retention. As shown in this table, Multi-R achieves the best speaker similarity MOS in terms of most emotion categories except for \textit{disgust}. However, this good target timbre retention is accompanied by unacceptably terrible emotion transfer as shown in Table~\ref{tab:compare_emomos}. Actually, the speaker similarity MOS difference between Multi-R and other methods could be partially caused by the fact that emotional speech could affect participants on the rating of the timbre similarity, because the reference is neutral rather than emotional speech. As for the comparison among CSET, PB, and the proposed CSPC, there is no obvious difference. Considering the obvious superiority of the proposed CSPC on the emotion transfer, the proposed method achieves new state-of-the-art performance on the \textit{cross-speaker} emotion transfer TTS task.

\subsection{Ablation study}
\label{sc:ablation}

In the proposed CSPC, emotion information is provided by embeddings with the emotion encoder in SDM and the prosody compensation encoder in PCM. In this section, we would like to show the necessity of utilizing these two embeddings on the emotion transfer. Furthermore, the performance of the GC block in the prosody compensation encoder would also be analyzed in an ablation study.

\begin{table}[h]
 \caption{Performance of different variants of the proposed methods in terms of emotion similarity with confidence intervals of 95\%.}
 \label{tab:ablation_emomos}
\setlength{\tabcolsep}{3mm}
 \centering
\begin{tabular}{l|l|l|l}
\toprule
\multicolumn{1}{c|}{Method} & \multicolumn{1}{c}{w/o EE} & \multicolumn{1}{c}{w/o PCE} & \multicolumn{1}{c}{CSPC} \\ \midrule
    surprise   &3.68$\pm$0.04 &3.77$\pm$0.04  &\bf{3.92}$\pm$0.06 \\
    happy     &3.75$\pm$0.05  &3.81$\pm$0.05  &\bf{4.01}$\pm$0.05 \\
    sadness   &3.44$\pm$0.04  &3.96$\pm$0.04  &\bf{4.05}$\pm$0.05 \\
    angry     &3.39$\pm$0.06  &3.72$\pm$0.04  &\bf{3.85}$\pm$0.07 \\
    disgust   &3.40$\pm$0.07  &3.76$\pm$0.05  &\bf{3.81}$\pm$0.06 \\
    fear      &3.42$\pm$0.04  &3.71$\pm$0.06  &\bf{3.98}$\pm$0.05 \\\midrule
    average  &3.51$\pm$0.04 &3.78$\pm$0.4  &\bf{3.89}$\pm$0.05 \\
  \bottomrule
\end{tabular}
\vspace{-0.1cm} 
\end{table}

\begin{table}[h]
 \caption{Performance of different variants of the proposed methods in terms of subjective speaker similarity with confidence intervals of 95\%.}
 \label{tab:ablation_spksorce}
\setlength{\tabcolsep}{3mm}
 \centering
\begin{tabular}{l|l|l|l}
\toprule
\multicolumn{1}{c|}{Method} & \multicolumn{1}{c}{w/o EE} & \multicolumn{1}{c}{w/o PCE} & \multicolumn{1}{c}{CSPC} \\ \midrule
    average &\bf{4.01$\pm$0.04} &3.96$\pm$0.061  &3.94$\pm$0.054 \\
  \bottomrule
\end{tabular}
\end{table}

\vspace{-0.2cm}
\subsubsection{The necessity of emotion embedding and prosody compensation}
\label{sc:ablation_1}
The verification of the necessity to use both emotion embedding and prosody compensation embedding is performed in an ablation study, in which the emotion embedding and prosody compensation embedding are dropped respectively. The subjective MOS test results are shown in Table~\ref{tab:ablation_emomos} and Table~\ref{tab:ablation_spksorce}, in which ``w/o EE" and ``w/o PCE” means the emotion embedding and prosody compensation embedding are dropped respectively. 

As shown in Table~\ref{tab:ablation_emomos}, the drop of either emotion embedding (w/o EE) or prosody compensation embedding (w/o PCE) results in an obvious performance drop in terms of all emotion categories. Specifically, the drop of emotion embedding brings the biggest performance drop on the emotion transfer, indicating that the emotion information is mainly conveyed by the emotion embedding. Even if the obvious performance drop appears when the emotion embedding is dropped, the performance of ``w/o EE" still outperforms Multi-R and PB (see Table~\ref{tab:compare_emomos}), demonstrating that the prosody compensation embedding indeed can convey the emotion information. The performance drop of ``w/o PCE" demonstrates that the prosody compensation embedding can provide extra emotion information to the emotion embedding. As for the target timbre retention, by comparing the speaker similarity MOS scores achieved by ``w/o PCE" and ``CSPC" in Table~\ref{tab:ablation_spksorce}, it can be found that no obvious performance drop appears when the prosody compensation embedding is added. All those results indicate that the ASR model’s intermediate feature (AIF) can effectively provide prosodic information to enable the proposed prosody compensation method to further improve the emotion transfer performance and meanwhile maintain the target speaker's timbre. 


\subsubsection{The effectiveness of GC block}
\label{sc:ablation_2}

The effectiveness of the emotion compensation embedding has been verified in Section~\ref{sc:ablation_1}. Here, we would like to show the importance of the GC block in the proposed emotion compensation encoder. To this end, an AB preference test between the proposed method and a variant of the proposed method, named as ``w/o GC", in which the GC block is dropped from the emotion compensation encoder (See Fig.~\ref{fig:frame2}) is conducted. 

\begin{figure}[ht]
\begin{minipage}[b]{1.0\linewidth}
  \centering
  \centerline{\includegraphics[height=4cm,width=8.5cm]{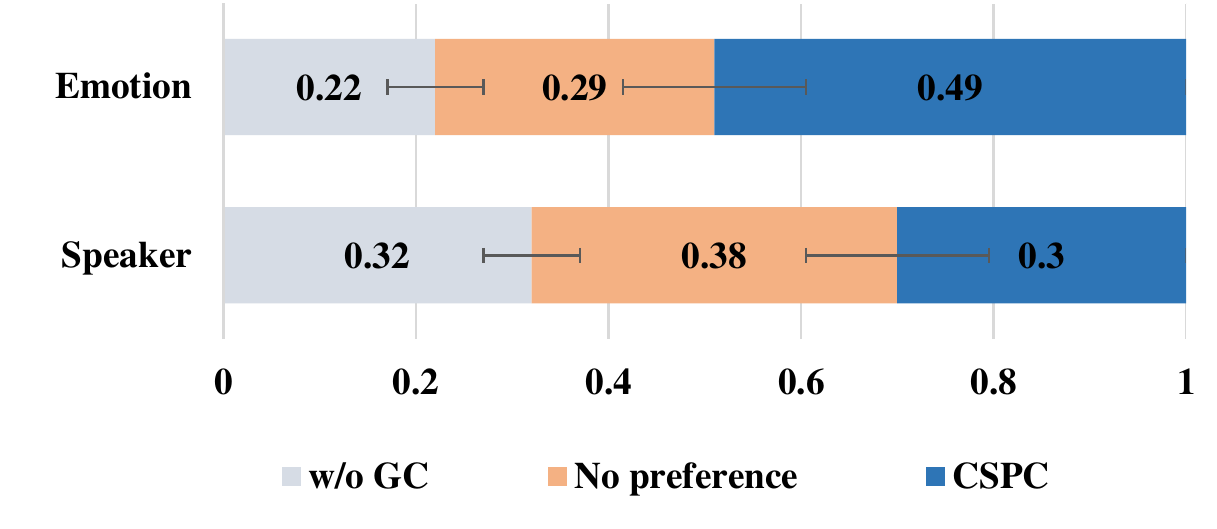}}
 \vspace{-0.2cm}
  \caption{Emotion similarity and speaker similarity AB preference test for the ablation study of GC block with confidence intervals of 95\%.}
  \label{fig:abtestemo}
 \end{minipage}
\vspace{-0.5cm} 
\end{figure}

The results are shown in Fig.~\ref{fig:abtestemo}, in which the global voting percents over different emotion categories are presented on the demo page\footnote{\label{demos}Audio samples and detailed results in terms of each emotion category can be found on the project page \url{https://silyfox.github.io/cspc/}} (Additional demos are also provided which transfer emotion to the new target speaker who only has 200 sentences through simple fine-tuning.). As can be seen from this figure, in terms of the emotion similarity, CSPC significantly outperforms the variant without the GC block (p-value = 0.042). As for the speaker similarity, dropping the GC block brings a slight increase which is not significant (p-value = 0.063). These results demonstrate that the proposed prosody compensation encoder is well designed, and benefits from the GC block on capturing the long-range dependency emotion information, which can better obtain the global emotion-related information from AIF.


\section{Conclusion}
\label{sec:illust}

In this paper, a prosody compensation method is proposed to improve the emotion transfer in the cross-speaker emotion transfer task. Inspired by the fact that the intermediate feature obtained by a pre-trained ASR model could contain prosody information, the prosody embedding extracted by the proposed prosody compensation module (PCM), which takes the ASR model's intermediate feature (AIF) as input, is used to provide extra information to the disentangled emotion embedding produced by a speaker disentangling module (SDM). Extensive experiments have demonstrated that the proposed prosody compensation method can successfully provide extra emotion information to the emotion embedding, for which the lost emotion information due to the speaker-emotion disentangling process could be well compensated. Furthermore, the ablation study shows that the GC block plays an important role in obtaining the prosody embedding, indicating the good design of the proposed prosody embedding encoder in PCM.

\vfill\pagebreak
\bibliographystyle{IEEEtran}

\bibliography{mybib}


\end{document}